\documentclass[final,5p,times,twocolumn]{elsarticle}

\usepackage{graphicx}
\usepackage{subfigure}
\usepackage{flushend} % balance cols on last page
\usepackage{ccicons}
 \usepackage{makerobust} 
 \MakeRobustCommand\rotatebox 
\graphicspath{{./pdf/}{./jpeg/}{./images/}}
 \DeclareGraphicsExtensions{.pdf,.jpeg,.jpg,.png}
 \DeclareGraphicsExtensions{.pdf,.jpeg,.jpg,.png} 
 \usepackage{amssymb}% http://ctan.org/pkg/amssymb 
 \usepackage{pifont}% http://ctan.org/pkg/pifont 
\usepackage{amssymb}
\usepackage[nodots,nocompress]{numcompress}

\usepackage{textcomp}

\usepackage{upgreek}

\usepackage{url}
\usepackage{float}

\usepackage{amsmath}

\usepackage[T1]{fontenc}

\usepackage{array}

\usepackage{color}

\usepackage{units}

\usepackage{soul}
\usepackage{color}
	\definecolor{celadon}{rgb}{0.67, 0.88, 0.69}
    \definecolor{flamingopink}{rgb}{0.99, 0.56, 0.67}

\usepackage{makecell}

\usepackage{enumitem}

\usepackage[table,xcdraw]{xcolor}
\usepackage{listings}
\biboptions{numbers}

\begin{document}

\begin{frontmatter}

%\title{Under-age Facial Age Recognition: Overview of Existing Techniques and Performance Evaluation}

%\title{Improving Automated Borderline Adult Facial\\Age Determination through Ensemble Learning}

%\title{To believe, or not to believe?: Forensic Insights from IoT through\\Electromagnetic Side-Channel Analysis}
%\title{To believe, or not to believe?: Forensic Insights from IoT through\\Electromagnetic Side-Channel Analysis}
%\title{Forensic Insights from IoT through Electromagnetic Side-Channel Analysis}
\title{Leveraging Electromagnetic Side-Channel Analysis for the Investigation of IoT Devices}

% removed the term CSAM because in the paper it's stated that the term CSEM would be used

%\author{Removed for blind review.}
%\ead{Removed for blind review.}
%\address{Removed for blind review.}

\author{Asanka Sayakkara~\corref{mycorrespondingauthor}}
\ead{asanka.sayakkara@ucdconnect.ie}
\author{Nhien-An Le-Khac}
\ead{an.lekhac@ucd.ie}
\author{Mark Scanlon}
\ead{mark.scanlon@ucd.ie}
\address{Forensics and Security Research Group, University College Dublin, Ireland}

\begin{abstract}

Internet of Things (IoT) devices have expanded the horizon of digital forensic investigations by providing a rich set of new evidence sources. IoT devices includes health implants, sports wearables, smart burglary alarms, smart thermostats, smart electrical appliances, and many more. Digital evidence from these IoT devices is often extracted from third party sources, e.g., paired smartphone applications or the devices' back-end cloud services. However vital digital evidence can still reside solely on the IoT device itself. The specifics of the IoT device's hardware is a black-box in many cases due to the lack of proven, established techniques to inspect IoT devices. This paper presents a novel methodology to inspect the internal software activities of IoT devices through their electromagnetic radiation emissions during live device investigation. When a running IoT device is identified at a crime scene, forensically important software activities can be revealed through an electromagnetic side-channel analysis (EM-SCA) attack. By using two representative IoT hardware platforms, this work demonstrates that cryptographic algorithms running on high-end IoT devices can be detected with over $82$\% accuracy, while minor software code differences in low-end IoT devices could be detected over $90$\% accuracy using a neural network-based classifier. Furthermore, it was experimentally demonstrated that malicious modification of the stock firmware of an IoT device can be detected through machine learning-assisted EM-SCA techniques. These techniques provide a new investigative vector for digital forensic investigators to inspect IoT devices.

\end{abstract}


\end{frontmatter}

%%%%%%%%%%%%%%%%%%%%%%%%%%%%%%%%%%%%%%%%%%%%%%%%%%%%%%%%%
\section{Introduction}
\label{intro}
%%%%%%%%%%%%%%%%%%%%%%%%%%%%%%%%%%%%%%%%%%%%%%%%%%%%%%%%%

The Internet of Things (IoT) has revolutionized the landscape of digital forensic investigations like never before. With the increasing prevalence of IoT devices in everyday life, these devices are capable of storing vital information that can prove useful in a digital investigation. A medical implant, such as a pacemaker, can provide hints in an investigation about a person of interest's physical exertion or stress introduced elevation of heart rate. A sports wearable, such as \emph{Fitbit}, can provide vital information about the presence and movements of a person in a crime scene. A smart smart voice assistant device, such as \emph{Amazon Alexa}, can provide a vital information about the time its owner came home. This kind of digital evidence is not available in traditional digital forensics, where the only resort was non-volatile storage of personal computers and removable media~\cite{chernyshev2018internet, quick2018iot, yaqoob2019internet}.

IoT devices are usually connected to the outside world in two main ways. Most IoT devices are connected to a cloud-based service through the Internet. This connection can either go directly to the cloud servers or in some cases delivered to a smartphone based app~\cite{lomotey2018traceability}. When an IoT device is subject to a digital forensic investigation, the digital evidence is often acquired from the associated smartphone app and/or the cloud servers as opposed to directly from the IoT device itself. Most IoT devices do not store a sufficient amount of data due to the limitation of local storage. Therefore, it is fair to look for IoT data in the user's smartphone or cloud storage. However, the reliability of digital evidence acquired from other places completely depends on the reliability of the IoT device itself. There is no guarantee that an IoT device is running the manufacturer's default firmware. If the device's firmware has been tampered with, all the digital evidence acquired from the associated smartphone app or the cloud servers may become unreliable.

Forensic inspection of IoT devices is a challenging task for digital forensic investigators. These devices lack common interfaces that can be used to acquire data using traditional forensic evidence gathering techniques~\cite{lillis2016challenges}. Most IoT devices follow proprietary hardware architectures and use low power consumption processors. Due to this reason, collecting forensically useful information directly from an IoT device often requires invasive techniques, such as tapping into the internal circuitry of the device or using chemicals to expose the silicon wafer of the flash data storage chips in order to extract data by physical means. For example, scanning electron microscopy (SEM) has been demonstrated to be useful in extracting data stored on EEPROM chips~\cite{torrance2009state, courbon2016reverse}. Such invasive approaches come with the risk of destroying or tampering the data stored on the target device. In order to perform forensic evidence gathering from IoT devices in a reliable manner, it is highly necessary to find non-invasive methods.

This work shows that unintentional electromagnetic (EM) radiation from IoT devices can be a potential non-invasive window to gather forensically useful information. The EM radiation patterns from the CPU of IoT devices sufficiently correlate to the software activities. Using a pair of \emph{Raspberry Pi} and \emph{Arduino Leonardo} devices as the general purpose IoT target device, this work shows that multiple forensically useful software behavior related information can be detected. Cryptographic algorithms running on a IoT device can be detected with $82$\% accuracy while variations of the software code behaviour can be detected with $90$\% accuracy through a combination of EM-SCA techniques and machine learning in a real-world setting.

\subsection{Contribution of this Work}
\label{contribution}

\begin{itemize}[noitemsep,topsep=0pt]
    \item As a solution to the challenge of extracting digital evidence from IoT devices using traditional approaches, this work introduces the potential of EM-SCA as a vector for gathering forensically useful insights from IoT devices.
    \item Experimentation and empirical evidence shows that the software behaviour of IoT devices can be reliably detected using machine learning techniques with over $80$\% accuracy through EM emissions in practical scenarios. This includes cryptographic algorithms that are employed to protect data stored on these devices.
    \item In order to integrate the EM-SCA techniques to gather forensically useful insights from IoT devices in practical digital forensic work flow, this work proposes a methodology for applying the techniques with minimum overhead and changes to existing digital forensic practices.
\end{itemize}

%%%%%%%%%%%%%%%%%%%%%%%%%%%%%%%%%%%%%%%%%%%%%%%%%%%%%%%%%
\section{Related Work}
\label{em-side-channels}
%%%%%%%%%%%%%%%%%%%%%%%%%%%%%%%%%%%%%%%%%%%%%%%%%%%%%%%%%

As derived from Maxwell's equations, EM waves can be generated by electrical currents varying over time. Characteristics of the EM waves being generated, such as frequency, amplitude, and phase, depends on the nature of the time varying electric current~\cite{maxwell1865dynamical}. Based on this principle, modern communication systems generate oscillating currents on antennas that generate EM waves that propagate over free space to be captured by another antenna with appropriate properties. The fact that modern digital computer systems have a large number of components that depend on electric pulses or alternating currents for their operations leaves the opportunity open for EM waves to be generated at unexpected frequencies without the intention of the system manufacturer~\cite{SAYAKKARA2019}.

In any computer, there are multiple components that operate in a coordinated, sequential fashion according to clock signals, including CPU~\cite{gandolfi2001electromagnetic}, RAM~\cite{gandolfi2001electromagnetic}, computer monitors~\cite{Sayakkara:2018:AEE:3230833.3234690}, etc. Among them, the CPU and RAM are most interest for the purpose of this paper. The CPU performs a cycle of fetching instructions, decoding them and executing them, while RAM maintains data and instructions when the device is powered on. The EM emission signals from these components contain a significant amount of side-channel information regarding the events related to software execution and data handling. On most IoT devices, the CPU and RAM are incorporated in the microcontroller (MCU) chips used on the boards.

Kocher et al.~were the first to introduce power consumption based side-channel attacks; both \emph{simple power analysis} (SPA) and \emph{differential power analysis} (DPA)~\cite{kocher1999differential}. SPA collects power consumption variation (in mA) over time with a high sample rate, such as 5 MHz. The authors showed that the waveform of the power consumption, when plotted against time, contained patterns that corresponded to the instructions of the data encryption standard (DES) cryptographic algorithm. If SPA can reveal the sequence of operations, it follows that this sequence depends on the data being handled by the algorithm (due to conditional branching). Designing code to minimize data dependent branching, which does not show characteristic power consumption patterns for specific operations, can prevent attackers from recognizing what is executing on the device~\cite{zankl2018side}.
DPA is a technique that can be custom tailored for specific encryption algorithms. Kocher et al. used the DPA technique against DES~\cite{kocher1999differential}. The technique was able to guess the encryption key accurately, given sufficient cipher texts and the power traces for those encryption operations. The authors state that they have used DPA to reverse engineer various unknown algorithms and protocols on devices. The authors indicate that it may be possible to automate this reverse engineering process. Kocher et al. hints that these techniques (i.e., SPA, DPA) might be also achievable usable with EM emissions.

Callan et al. introduced a metric called SAVAT (Signal AVailability for an ATtacker) that measures the EM signal power emitted when a CPU is executing a specific pair of instructions (A and B). The authors show that different selections of A/B instruction pairs emit different SAVAT values, i.e., signal power~\cite{callan2014practical, zajic2014experimental}. An improvement to the SAVAT technique is a method called \emph{Finding Amplitude-modulated Side-channel Emanations} (FASE). The key premise behind the FASE technique relies on the phenomena that when a program activity is alternating at a frequency ($f_{alt}$) that affects any periodic EM signal originating from any source at a frequency $f_{c}$, it is possible to observe two side-band signals at $f_c - f_{alt}$ and $f_c + f_{alt}$ between the $f_c$ signal. Further improvements to the SAVAT technique enabled the possibility of identifying both amplitude and frequency modulated EM emissions from CPUs~\cite{callan2015fase, prvulovic2017method, yilmaz2018capacity}. While it is evident from existing studies that the EM side-channel leakage is available across various type of CPUs, further work is necessary to identify the effect of different CPU architectures to the EM emissions.

The simplest form of representing EM side-channel emission data is the waveform of the signal in the time domain. Stone et al.~built \emph{matched-filter classifiers} that utilize correlation between known EM signal waveform vectors with an unknown EM signal waveform vectors to detect software activities on microcontrollers used in embedded devices~\cite{stone2015radio}. In this work, the software activities considered were individual CPU operations such as \texttt{mov}, \texttt{add} and \texttt{sub} instructions. In order to generate matched-filter templates, an assembly program that executed a particular CPU operation continuously was used which triggers a general purpose input/output (GPIO) pin value. This GPIO trigger was separately probed in order to identify the boundaries of EM emission signals in order to extract the matched-filter template trace.

Stone et al.~continued to demonstrated that instead of using the time-domain signal as a feature vector, it is more effective to use \emph{Hilbert transformation} of the EM emission signals~\cite{stone2015detecting}. The advantage of this approach is that, when calculating correlation of two signals, Hilbert-transformed vectors perform better than time-domain vectors for the same signal-to-noise ratio (SNR) of signals. As in their previous work with time-domain signals, templates were generated for individual instructions running on a target device and these ware used with a correlation-based classifier to detect when a device was executing anomalous codes.

In addition to the time-domain signals and Hilbert-transformed signals, another alternative format of representing EM emission signals is using a RF-DNA fingerprint. RF-DNA fingerprinting is a technique to fingerprint the radio signals transmitted by various devices including WiFi, Bluetooth, Zigbee, GSM devices, RADAR antennas, etc. This technique has been used to identify rogue devices in a deployment through using their RF signals without physically inspecting them~\cite{reising2012exploitation, dubendorfer2013using, danev2012physical, lukacs2015rf}. Deppensmith et al. showed that the RF-DNA technique can be reliably applied to unintentional EM emission fingerprinting on computing devices~\cite{deppensmith2014optimized}. Lukacs et al.~used \emph{multiple discriminant analysis} (MDL) in order to reduce the dimensionality of RF-DNA fingerprints before applying them into a \emph{maximum likelihood} (ML) classifier to identify known radio transmitters used in radar systems~\cite{lukacs2015rf}. Similarly, Bihl et al.~showed that MDL can help in identifying most important features from RF-DNA fingerprints~\cite{bihl2016feature}.
However, the evaluations performed by Stone et al. on microcontroller based IoT devices indicates that further study is necessary to conclude the most reliable format to represent unintentional EM signals~\cite{stone2016comparison}.

Wang et al.~evaluated the possibility of using Multi-Layer Perceptron (MLP) and Long Short-Term Memory (LSTM) to detect software activities and modifications to the software through the changes in EM emissions~\cite{wang2018deep}. While they also used \emph{Arduino} and \emph{Raspberry Pi} devices as in this work, their evaluation was limited to a classification with fewer classes, i.e., 2 classes for \emph{Arduino}, 2 classes for \emph{Raspberry Pi}, and 5 classes for a programmable logic controller (PLC) device. In contrast, this work shows that it is possible to detect a wide variation of changes to a target IoT device by utilizing simpler machine learning models. Furthermore, Wang et al. depended on an instrumented \emph{Arduino} device in order to trigger the sampling device while the technique used as part of this paper can observe the \emph{Arduino} device without any instrumentation.

As existing work has shown, software activities running on IoT devices can be detected through EM-SCA techniques. In order to use these attacks for digital forensic purposes, these attacks must work in real-world conditions, such as target devices with zero or minimal knowledge, and devices placed in noisy environments~\cite{SAYAKKARA2019}.

%%%%%%%%%%%%%%%%%%%%%%%%%%%%%%%%%%%%%%%%%%%%%%%%%%%%%%%%%
\section{Electromagnetic Analysis on IoT Forensics}
\label{em-analysis-on-iot-forensics}
%%%%%%%%%%%%%%%%%%%%%%%%%%%%%%%%%%%%%%%%%%%%%%%%%%%%%%%%%

\begin{figure}[t!] %[!htbp]
\centering
\includegraphics[width=0.35\textwidth]{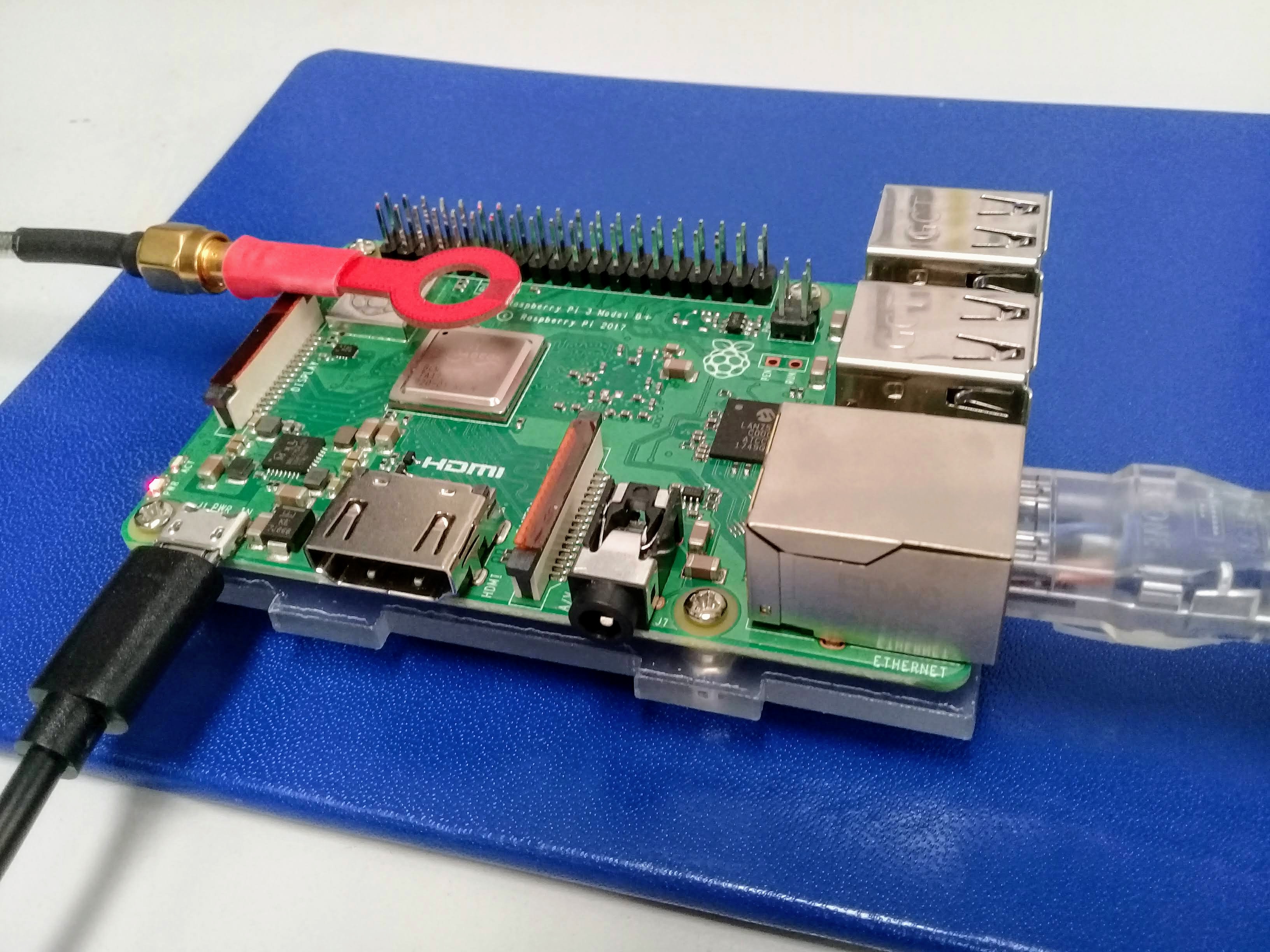}
\caption{EM emissions from Raspberry Pi are captured using an H-probe antenna placed closer to the processor.}
\label{fig:raspberrypi_with_antenna}
\end{figure}

\begin{figure}[t!] %[!htbp]
\centering
\includegraphics[width=0.45\textwidth]{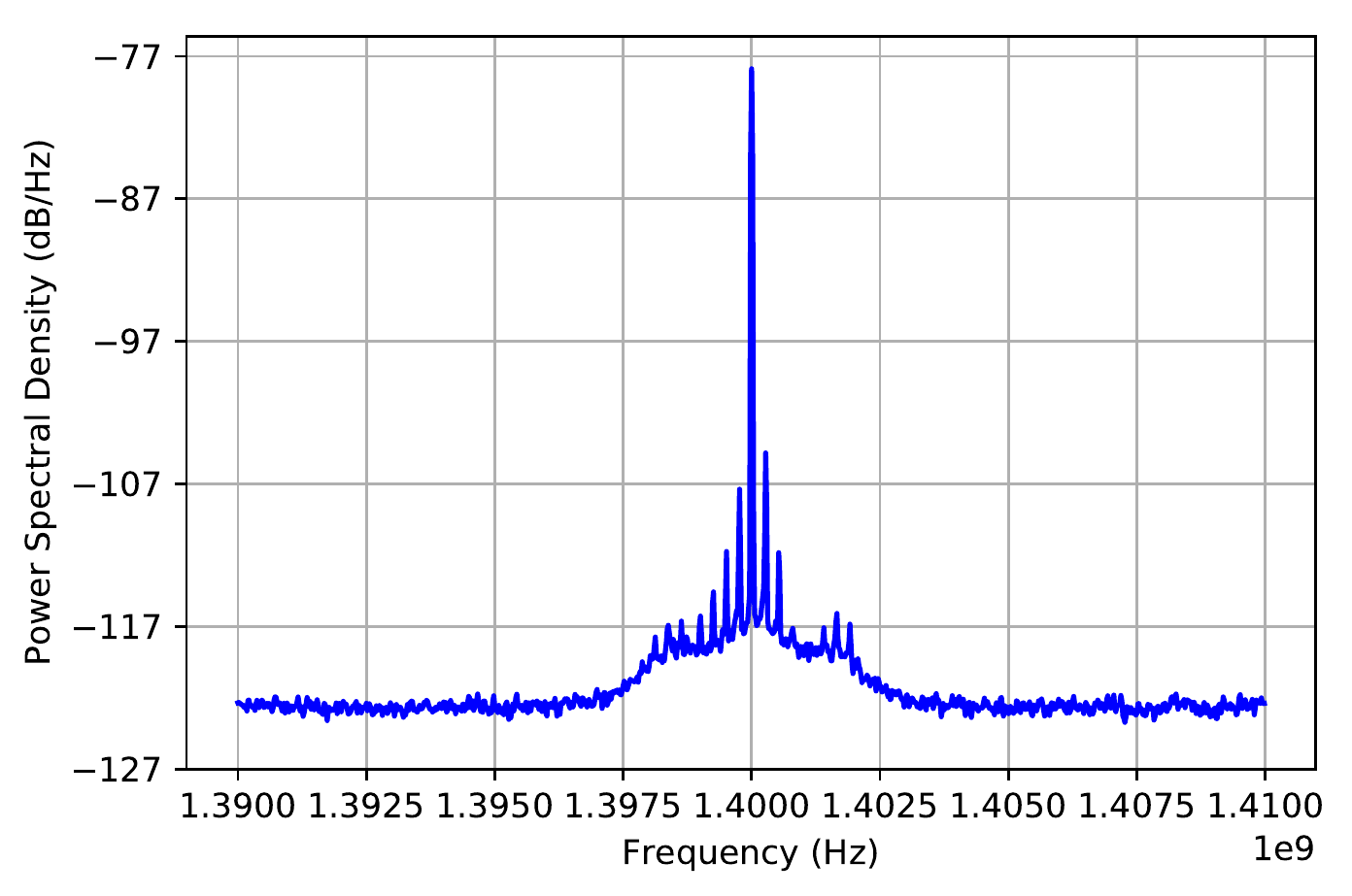}
\caption{Power spectral density (PSD) of the Raspberry Pi device as observed around the clock frequency. A sampling duration of $0.01$ seconds was used to capture EM signals for this work.}
\label{fig:rpi-psd}
\end{figure}

EM radiation can be generated from various components of an IoT device including the processor, network controller chips (both wired and wireless), video displays, sensors, actuators and many more. Among these EM noise causing components, the processor is the most significant component from a forensic point of view as it has been shown that the EM emission patterns of CPU can be correlated with the device's software activities. While any IoT device can be designed with a unique processor, there is an important commonality of components. There are only few common architectures used for microcontrollers in most IoT devices, e.g., ARM, AVR, and MSP430. This means, EM emission patterns identified from a particular processor chip should be applicable across many IoT devices that employ them.

Throughout the experimental study of this work, two representative devices were used; namely a \emph{Raspberry Pi 3 B+} and an \emph{Arduino Leonardo}. The \emph{Raspberry Pi 3 B+} device consists of a \emph{ARM Cortex-A53} quad-core processor running at $1.4$~GHz clock speed. It has a memory capacity of $1$~GB. Furthermore, it has WiFi, Bluetooth 4.0, and Ethernet for communication. All of these resources represent the class of a high-end IoT device that is capable of running a \emph{Linux}-based operating system and comparatively heavier applications. Therefore, this device can be used to easily emulate various existing IoT devices during experimentation. Meanwhile, the \emph{Arduino Leonardo} device consists of an $8$~bit micro-controller with an AVR architecture that runs at $16$~MHz clock speed. It has $2.5$~KB memory which is barely enough to run simpler programs. Therefore, it can be considered as a representative of a lower-end IoT device.

This section demonstrates the use of EM-SCA techniques in order to gather forensically useful insights from these target devices. First, the minimum required hardware setup to observe EM emissions from the target devices is introduced. Secondly, the details of multiple experiments is provided in order to identify various forensically useful insights from the target device, such as cryptographic operations and software behaviour related activities. Finally, it is demonstrated that software activity detection can be performed in real-time, making live forensic analysis of IoT devices through EM-SCA feasible.

The source code and machine learning models used in the experiments of this paper are available at a \emph{Github} repository\footnote{https://github.com/asanka-code/dfrws-2019-code} for reproduction of the results.

%----------------------------------
\subsection{Observing EM Emissions}

In order to acquire EM emissions from the target device, a software defined radio (SDR) device called \emph{HackRF} was used~\cite{ossmann2016software}. In contrast to traditional hardware radios, SDR devices consist of minimal hardware components with most of the signal processing performed using software. Therefore, SDR devices are highly flexible and easy to use~\cite{tuttlebee2003software}. The \emph{HackRF} device has a maximum sampling rate of $20$~MHz, where each sample has a $8$~bit resolution. The device can be tuned to a wide range of frequencies between $1$~MHz to $6$~GHz. The device contains built-in receiver amplifiers in order to enhance the captured EM signals before digitizing them. While various types of antennas can be connected to this SDR device, small H-loop antennas are the most appropriate for the purpose of capturing EM radiation coming from small components of IoT devices, such as processors.

When performing experiments to observe EM emissions from the Raspberry Pi, it was connected to a host computer through the Ethernet port and logged into remotely through a secure shell (SSH). This enables the control of the Raspberry Pi through the host computer remotely during experimentation. The SDR device is connected to the same host computer to store the captured EM data. As shown in Figure~\ref{fig:raspberrypi_with_antenna}, the H-loop antenna of the SDR is placed right on top of the processor chip of the Raspberry Pi leaving a gap of approximately $1$~cm in order to maximize the reception of EM emissions. While the experimental setup keeps the signal acquisition antenna closer to the target device, it is possible to use directional antennas and signal amplifiers to observe EM emissions from IoT devices at large distances up to several meters~\cite{juyal2018directive}.

In order to observe EM emissions from a target device's CPU, the EM emission frequency needs to be determined. The clock frequency of the CPU is the most fundamental frequency for EM radiation. Furthermore, harmonics of this fundamental frequency can also contain the desired information. Therefore, the exact choice of the frequency depends on what has the highest amplitude with the least amount of external interference.
The processor of the Raspberry Pi emits EM radiation at several different frequencies and their associated harmonics. The most reliable frequency observed from the device was the fundamental clock frequency of the processor, which is $1.4$~GHz. Figure~\ref{fig:rpi-psd} illustrates the power spectral density (PSD) of the EM emissions surrounding this specific clock frequency. A strong peak at $1.4$~GHz is evident alongside multiple side-bands.

\begin{figure}[t!] %[!htbp]
\centering
\includegraphics[width=0.5\textwidth]{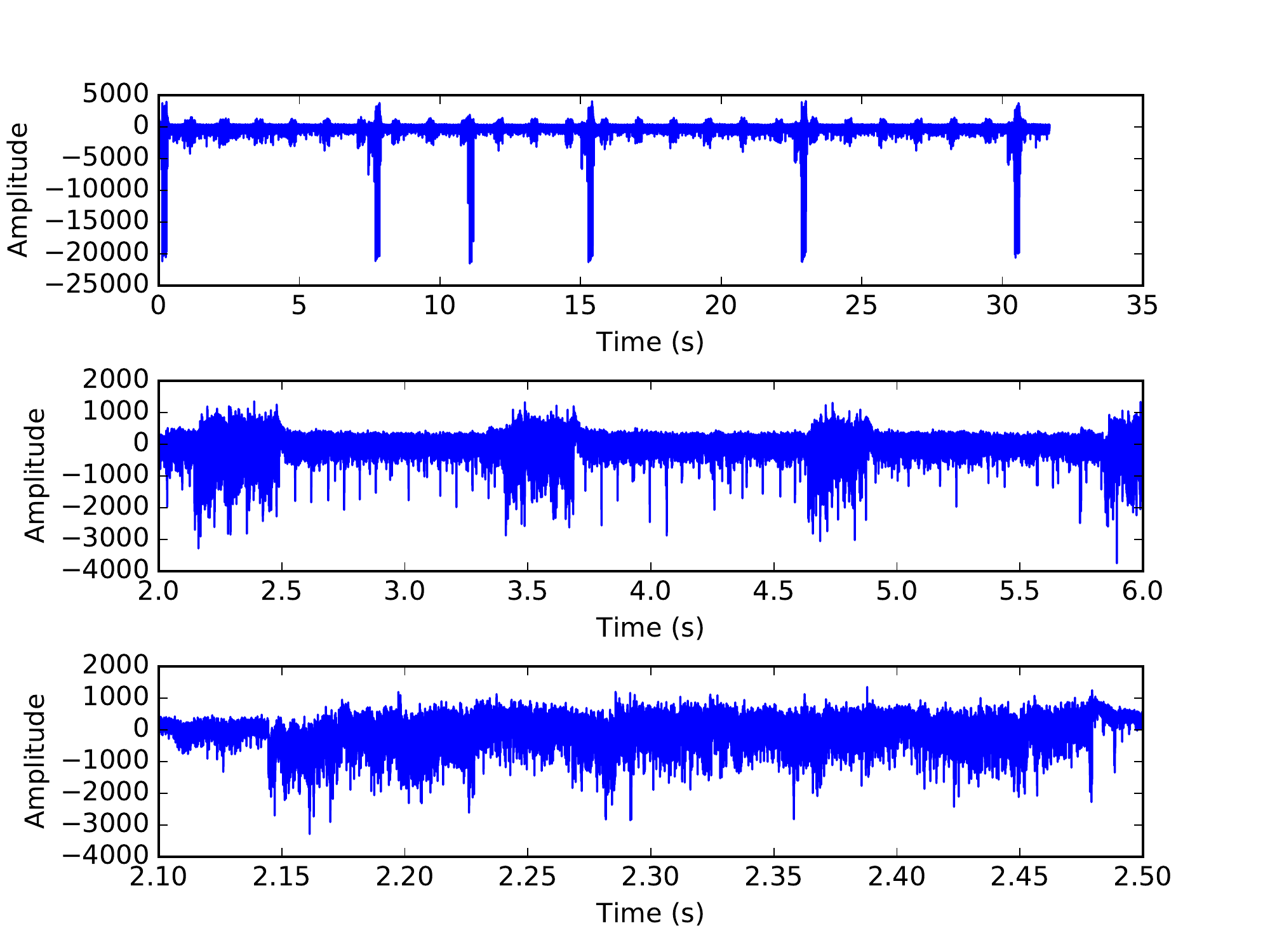}
\caption{Waveform of the AM demodulated signal at the CPU clock frequency of Raspberry Pi. The AM modulated signal represents the AES encryption performed on the device. Sudden higher peaks are an external interference signal coming from an unknown source.}
\label{fig:rpi-am-demod}
\end{figure}

\begin{figure}[t!] %[!htbp]
\centering
\includegraphics[width=0.47\textwidth]{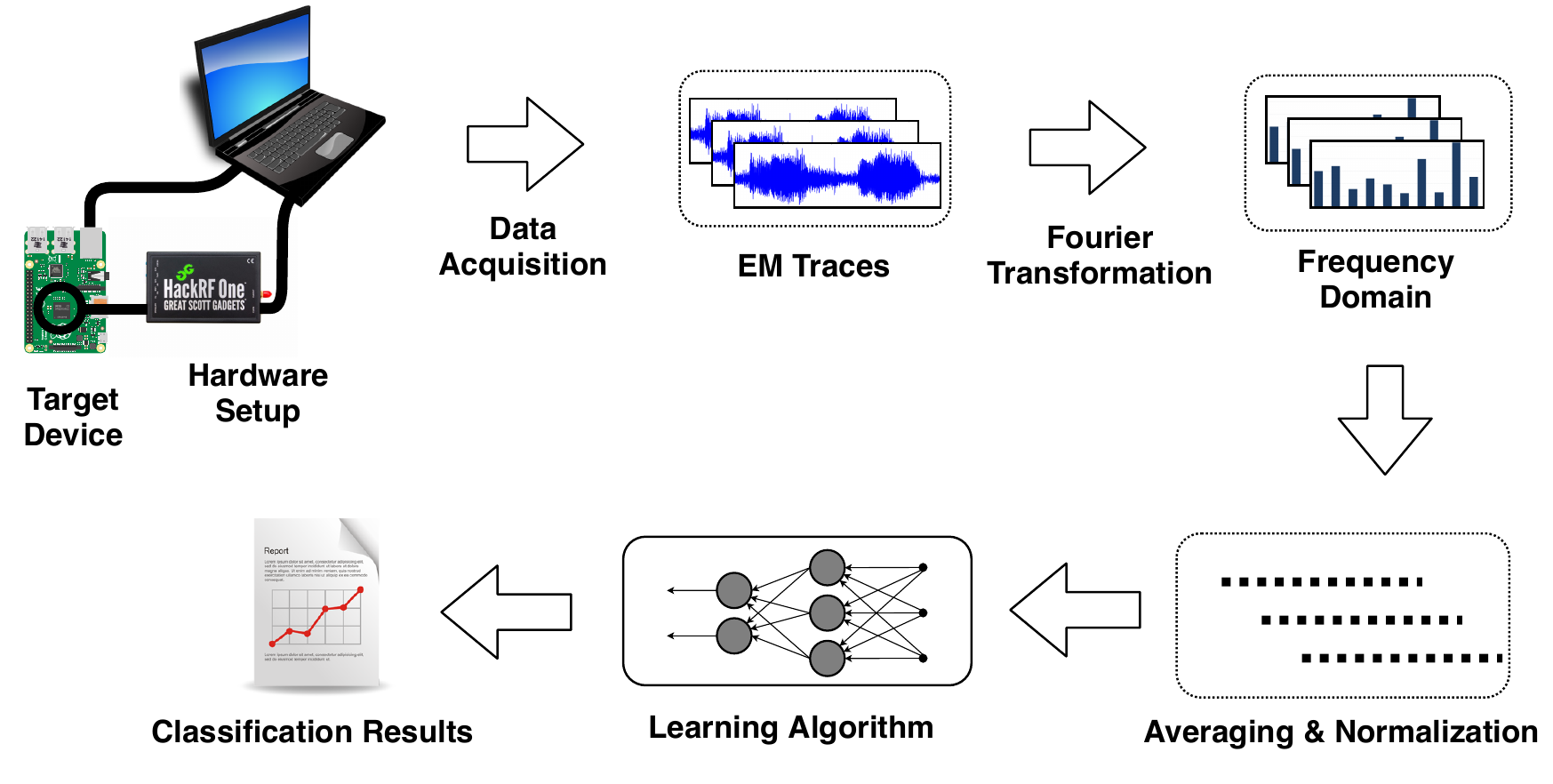}
\caption{The EM trace acquisition and preprocessing stages in order to classify cryptographic activities using a machine learning model.}
\label{fig:crypto-classifier-steps}
\end{figure}

As previously shown in the literature~\cite{callan2014practical, zajic2014experimental}, EM signals coming from the CPU modulates the software behaviour on its amplitude. In order to observe such variations, following experiment was performed. The Raspberry Pi was programmed to run a shell script that performed AES encryption operations with a time gap of $1$~second. The shell script used \emph{OpenSSL} commands to invoke the \emph{AES-256-CBC} algorithm on a large file continuously. The AES operations performed periodically on the Raspberry Pi resulted in observations of amplitude variations in the EM signal, as illustrated in Figure~\ref{fig:rpi-am-demod}. The blobs that occur with a 1 second gap in the first sub-figure correspond to the AES encryption operations, while the much higher peaks that occur at irregular intervals are external noise. A selected region of the signal is zoomed-in in the second sub-figure. The third sub-figure illustrates the EM emission pattern of a single AES encryption operation.

When observing EM emissions from the Arduino Leonardo, the clock frequency of $16$~MHz becomes the first potential target. However, through several empirical observations, it was identified that some higher harmonics leak more information than fundamental clock frequency of the device. Among them $288$~MHz, which is the $18^{th}$ harmonic of the clock frequency. This frequency was selected to be the information leaking channel in the experimentation presented as part of this work. 

\subsection{Discriminating Cryptographic Activities}

Among the software activities of IoT devices that may have a forensic interest, cryptographic operations are at the forefront. When storing data on-board or transmitting over the network, modern high-end IoT devices tend to rely on cryptographic encryption as a security measure. The following experiment investigates the possibility of using EM emissions of IoT devices in order to automatically detect when they perform data encryption operations. Three major cryptographic algorithms, i.e., AES-128, AES-256, and 3DES are used, as three classes and a mixture of non-cryptographic operations is used as another class. Figure~\ref{fig:crypto-classifier-steps} illustrates the procedure of acquiring data, preprocessing data, and finally the classification to classes.

\textbf{Data Acquisition:} The same hardware configuration with a Raspberry Pi as the target device and a HackRF as the EM signal capturing device are used in this experiment. In order to train a classifier to detect cryptographic operations, a labeled EM trace set is necessary for each classification class. To assist in this task, a UDP communication channel between the Raspberry Pi and the host computer was established through the Ethernet cable. To reduce unnecessary EM noise capture, each time the Raspberry Pi perform a cryptographic operation, it notified the host computer immediately before and after by sending UDP packets. This allows the host computer to identify the time period of the EM data stream coming from the HackRF, that corresponds to the cryptographic operation. Future work will focus on automatically identifying the necessary emissions through a sliding window, eliminating this step. Each such identified EM signal segments are saved as an EM trace along with the label of the cryptographic algorithm. Overall, the EM traces collected was about $12$~GB.

\textbf{Data Preprocessing:} Due to multiple reasons, the acquired EM can have variable lengths in the time-domain and also may not properly enclose the cryptographic operation within its boundary. These reasons include the inherent difference of the time each cryptographic calculation takes to execute, the delays in UDP communication between the Raspberry Pi and the host computer, and the delays in the HackRF data acquisition software to start and stop the EM sampling. Due to the large length and the variability of the lengths, these labeled EM trace samples are still unsuitable to be directly used as training samples for a machine learning-based classifier. To mitigate these differences in EM traces, each trace is converted into the frequency-domain by using a \emph{Fourier Transformation}. This is achieved by taking a segment of $0.1$~seconds from the beginning of each EM trace and applying Fast Fourier Transform (FFT)~\cite{smith1997scientist}.

\begin{figure}[t!] %[!htbp]
\centering
\includegraphics[width=0.5\textwidth]{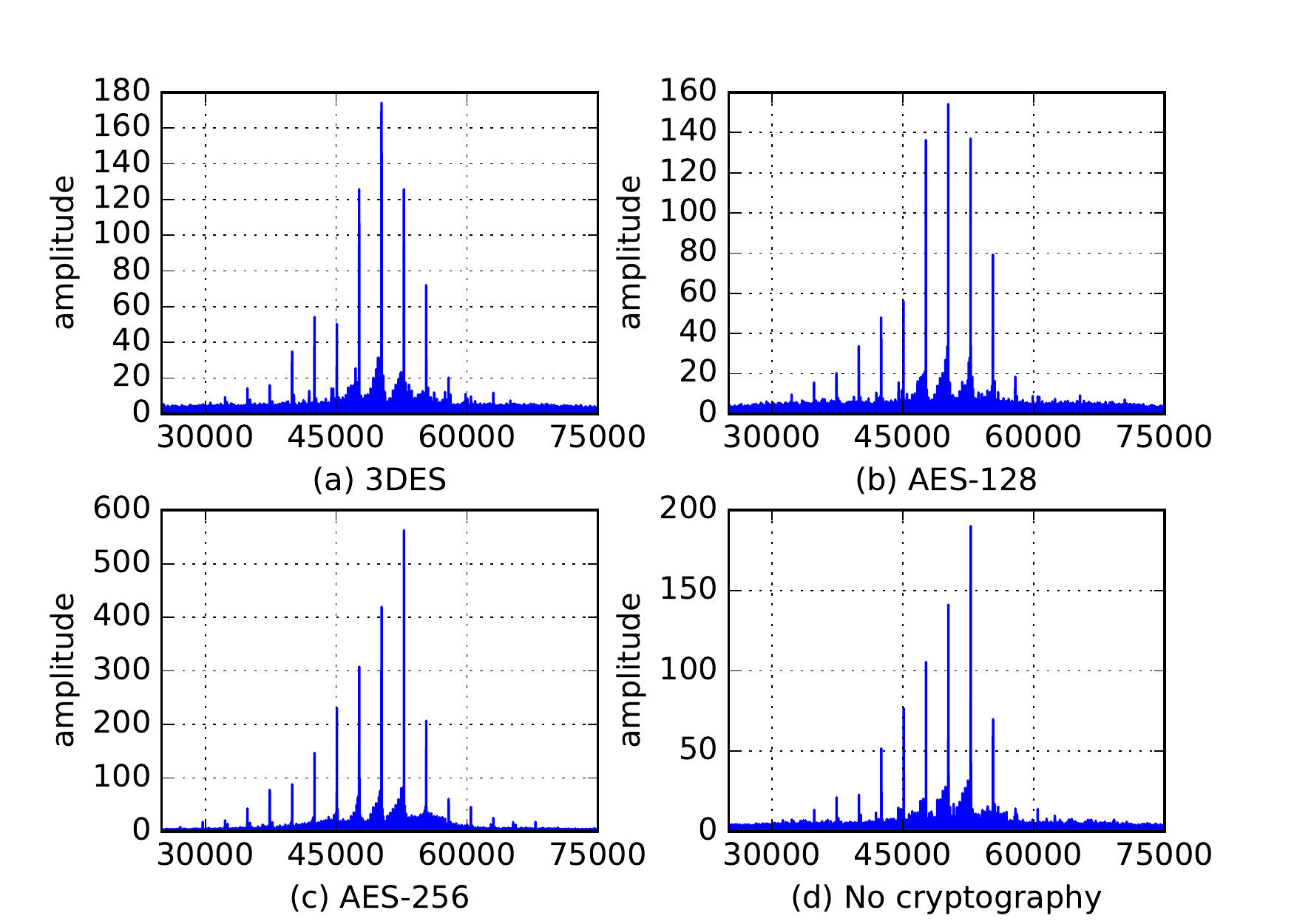}
\caption{Samples Fourier Transform vectors of cryptographic algorithms that run on Raspberry Pi.}
\label{fig:fft-plots-of-neural-network-classification}
\end{figure}

Since the sampling rate of the HackRF is $20$~MHz, the resulting Fourier Transform contained a vector with 200,000 elements; each containing the amplitude of a frequency component of the original EM emission. In this Fourier Transform vector, it was observed that the variation of peaks from software activity was only distinguishable at the middle portion. Therefore, it was decided to use only the frequency components from \nicefrac{1}{4} to \nicefrac{3}{4} of the original Fourier Transform through discarding the edges. Figure~\ref{fig:fft-plots-of-neural-network-classification} illustrates samples of Fourier Transforms from each class, where it is evident that there are slight variations unique to each activity.

The number of elements in the Fourier Transform was too large to be directly taken as an input vector for modelling. The dimensions can be reduced by breaking the Fourier Transform vector into a limited number of buckets. Subsequently, a representative value can be selected for each bucket by averaging the values or selecting the maximum valued element in each bucket. In this particular experiment, $500$ buckets were selected where the elements within each bucket were averaged to generate feature vector of $500$ features. The number of buckets and feature vectors was decided through experimentation and evaluation of the produced machine learning classification models.

%From this selected range of the Fourier Transform, a feature vector of $500$ features was created by breaking the Fourier Transform vector into $500$ buckets and averaging the values within each bucket.

%\todo[inline]{@Asanka: "A neural network with two hidden layers". I think you also need to explain a little bit why you chose 10 and 5 nodes for these two hidden layers. For example, you can say: we tried with different numbers of hidden nodes and these ones are the best.}

\begin{table}[t!] %[!htbp]
\centering
\begin{tabular}{|c|c|c|c|}
\hline
%\rowcolor[HTML]{DAE8FC} 
\textbf{Activity} & \textbf{Precision} & \textbf{Recall} & \textbf{F1-Score} \\ \hline
\textbf{Other}    & 0.93               & 0.85            & 0.89              \\ \hline
\textbf{AES-256}  & 0.78               & 0.86            & 0.82              \\ \hline
\textbf{AES-128}  & 0.99               & 0.92            & 0.95              \\ \hline
\textbf{3DES}     & 0.81               & 0.85            & 0.83              \\ \hline
\end{tabular}
\caption{Classification accuracy of cryptographic algorithms.}
\label{tab:crypto-classification-results}
\end{table}

\textbf{Classification:} A neural network was implemented to classify EM traces into the correct class that had four layers; an input layer, two hidden layers, and an output layer. The number of hidden layers and the number of hidden nodes used in each of the hidden layers were decided empirically by evaluating various settings. Accordingly, the first hidden layer was assigned 10 hidden nodes, while the second hidden layer was assigned 5 hidden nodes. The input layer has 500 input nodes for the feature vector and the output layer has 4 nodes for the four classes. For each class, $600$~samples were taken for the training process totalling 2,400~training samples for all four classes. The learning rate of the neural network was set to $1^{-20}$, which was decided empirically. The classifier code was running on a computer with $64$~bit \emph{Intel Core i-5 quad-core} processor and $16$~GB memory, running a Linux operating system. While the EM traces acquisition and preprocessing to generate training samples took several hours, the training and testing phases of the neural network took less than a minute to provide classification results. A 10-fold cross-validation was used to validate the classification results.

\textbf{Results:} The results of the classification is illustrated in Table~\ref{tab:crypto-classification-results}. The neural network classifier correctly classified the three cryptographic algorithms and the non-cryptographic scenarios with $80$\% accurately. Considering the fact that Raspberry Pi was running a computationally heavy operating system like \emph{Linux}, which can make use of all four cores of the processor for multi-tasking, the ability to distinguish between these three major encryption algorithm settings hints that it should be possible to detect cryptographic algorithms on much less capable hardware devices. Existing cryptographic key recovery attacks depend on prior knowledge of the cryptographic algorithm being employed. The ability to identify the cryptographic algorithm solely based on EM observations can increase the likelihood of success for such key recovery attacks.

%----------------------------------
\subsection{Detection of Software Behaviour}

While heavy cryptographic algorithms are employed on resource rich IoT devices, simpler devices are unable to use such computationally heavy algorithms to encrypt data due to the lack of computational resources. Therefore, they are usually programmed to perform a repetitive task continuously. Among the various tasks performed by IoT devices, certain tasks have forensic interest. These include reading data from a specific on-board sensors, such as a microphone, writing data to an on-board storage device, such as an SD card, and executing a command received remotely through the network. Identifying what operations an IoT device is performing at the moment when it was seized live could prove important. For example, if the device is currently wiping the SD card according to a command received remotely, the investigators need to know it immediately so that they can turn the device off without waiting for any further live analysis.

\begin{figure}[!t]
% \begin{minted}[mathescape,linenos,numbersep=4pt,frame=lines,framesep=2mm, samepage=true]{java}
% /* Arduino test program */
% void setup(){
% }
% void loop(){
%     for(int i=0, i<20, i++) { delay(10); }
%     for(int i=0, i<20, i++) { delay(10); }
%     /* further loops */
% }
% \end{minted}
\centering
\includegraphics[width=0.5\textwidth]{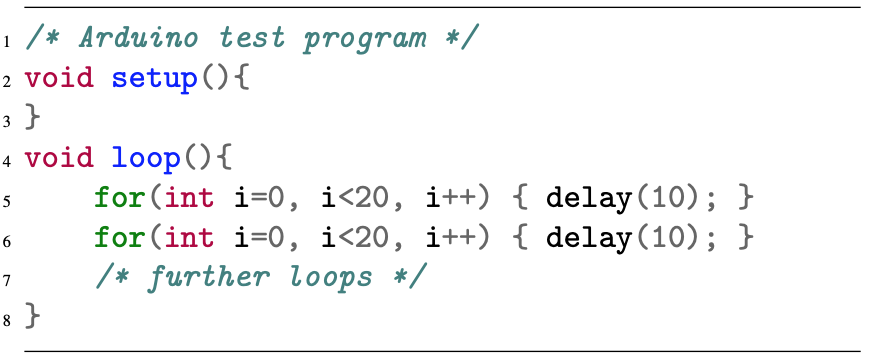}
\caption{An example \emph{Arduino} program which performs a time complexity $O(n)$ task repetitively inside an infinite loop which is used as a classification target.}
\label{lst:arduino-code}
\end{figure}

To explore the possibility of distinguishing different tasks performed by a simple IoT device, the following experiment was carried out. The objective was to train and test a machine learning model that can classify simple IoT firmware with increasing complexity. It was decided to use a Arduino device for this experiment as its simpler processor matches the resource profile of a lower-end IoT device. In order for classification, ten \emph{Arduino} programs were selected that repeatedly perform a task inside an infinite loop. Listing~\ref{lst:arduino-code} illustrates an example \emph{Arduino} program used as a classification target.
As can be seen, the program consists of an infinite loop designed to represent a repetitive task of an IoT device with a time complexity of $O(n)$. Each subtask the device is performing is represented by individual \texttt{for} loops with a finite number of iterations. It is assumed that a malicious modification to the device is performed by adding a new subtask to the program or by removing an existing subtask from the program.

\begin{figure}[!t] %[!htbp]
\centering
\includegraphics[width=0.5\textwidth]{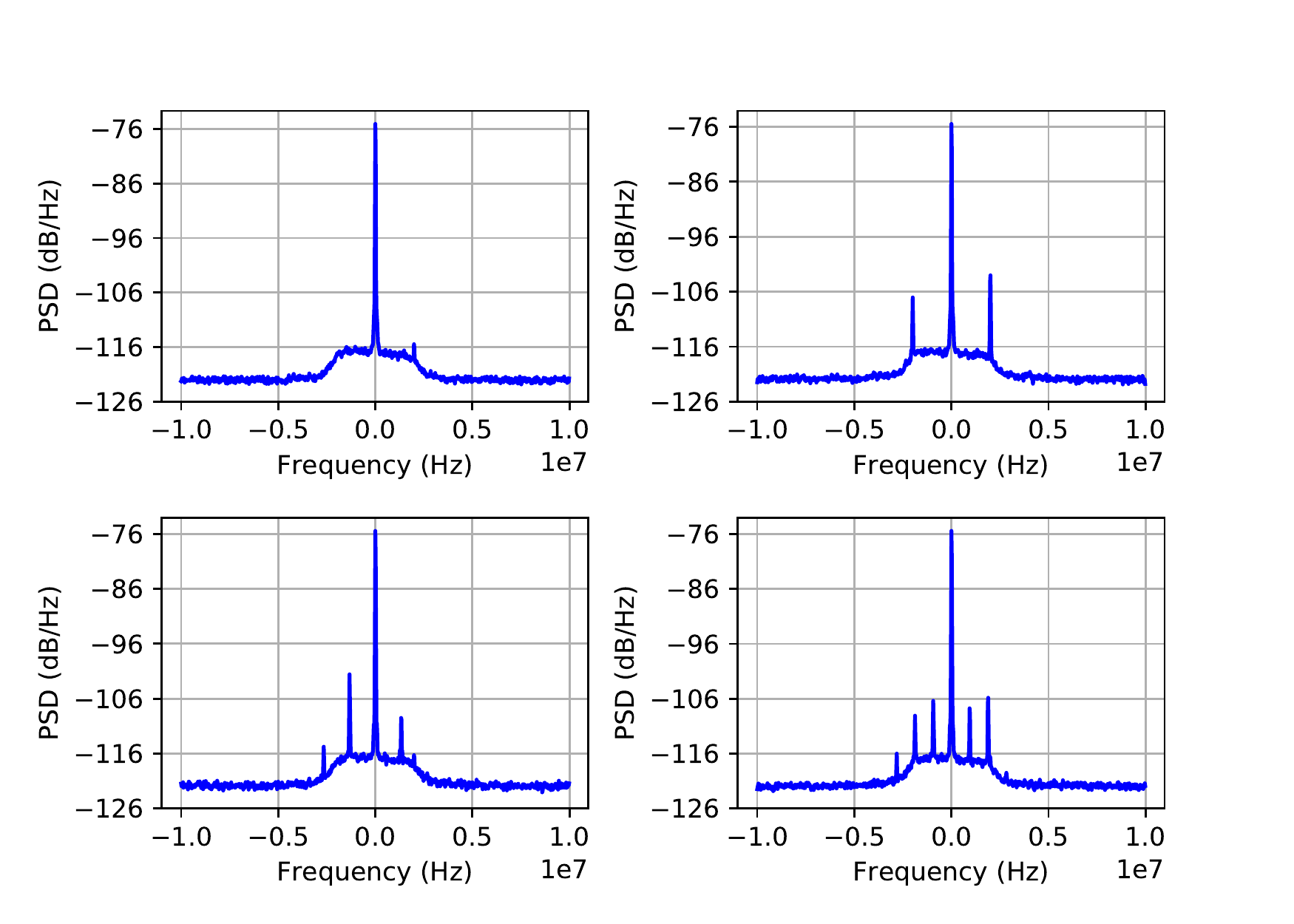}
\caption{Power spectral density (PSD) of the EM emission from four different \emph{Arduino} programs which were used for classification.}
\label{fig:arduino-psd-plots}
\end{figure}

\textbf{Data Acquisition:} In order to collect EM trace samples for each program, the Arduino was programmed with them separately and allowed to run with a H-loop antenna placed approximately $1$~cm above the microcontroller of the device. The HackRF was tuned to the information leaking $288$~MHz frequency of the target device and sampled data at the rate of $20$~MHz. Since the target device was performing a repetitive task, there was no software instrumentation required. Each acquired EM trace was approximately $25$~milliseconds long. Since there were ten programs to detect, $600$ EM traces were acquired per class, which resulted in $177$~GB of data for the overall 6,000 EM traces. Figure~\ref{fig:arduino-psd-plots} illustrates the power spectral density (PSD) of the EM emissions of four such programs subject to the experiment.

\textbf{Data Preprocessing:} From the extracted EM traces of each program class, $10$~milliseconds long segments were extracted and converted to the frequency domain using Fast Fourier Transformation (FFT). Unlike the aforementioned scenario of classifying between 4 cryptographic classes, this experiment attempts to classify 10 different programs. A 500 element feature vector did not seem to be effective in this case. Therefore, it was empirically decided to create a feature vector of 1,000 features by breaking a Fourier Transform vector into 1,000~buckets. Furthermore, it was noticed that averaging values within a bucket smoothed out the most significant frequency component under the noise floor. This most significant frequency ideally would have been selected as the representative element for the bucket. Therefore, it was decided to select the maximum value within each bucket instead of averaging in order to build the feature vector.

Even though the EM trace data were acquired while the target device was running in a noisy environment, there was no noise filtering applied to the EM traces before generating the feature vectors. The choice of the information leaking $18^{th}$ harmonic of the Arduino clock frequency was made to ensure no strong external noise source in that frequency.

%Explain how we converted raw EM traces into the feature vectors. From each EM trace, we extract a $10$~millisecond segment and convert to the frequency domain using Fast Fourier Transformation (FFT). Unlike with the \emph{Raspberry Pi}, we are using the whole Fourier Transform vector to generate the feature vector for machine learning algorithm. A feature vector of $1000$ features was generated by breaking a Fourier Transform vector into $1000$ buckets and then taking the maximum value of the bucket as a feature.  

\textbf{Classification:} Similar to the previous experiment, a neural network with two hidden layers was designed, where first hidden layer contained $10$ hidden nodes while the second hidden layer contained $3$ hidden nodes. The input layer contained 1,000 features and the output layer contains $10$ output nodes. With $600$ training samples for each class, a total of 6,000 training samples were fed to the neural network to train and test the model to detect ten Arduino programs running on the target device.

\begin{figure}[!t] %[!htbp]
\centering
\includegraphics[width=0.5\textwidth]{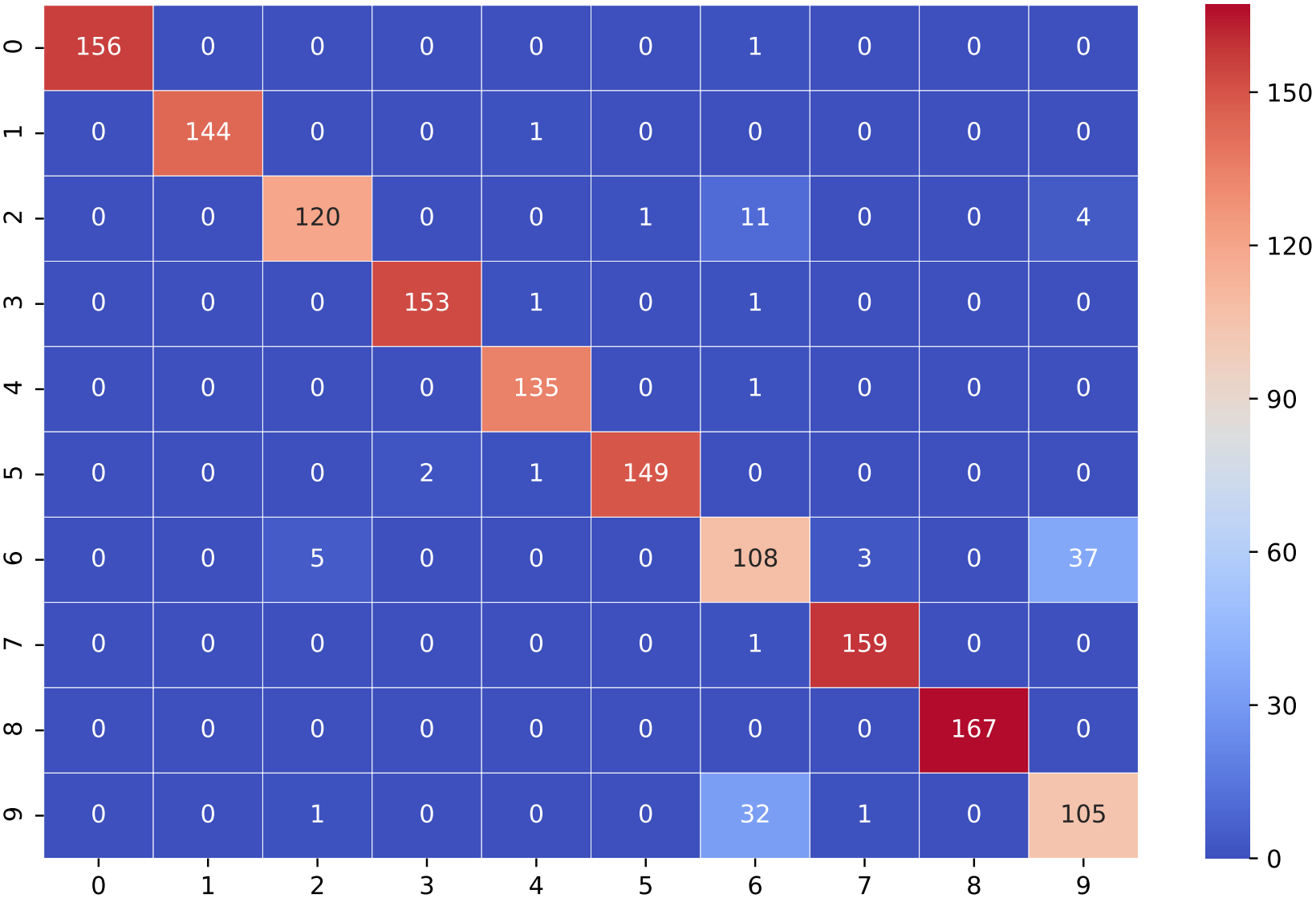}
\caption{Confusion matrix of the neural network classifier to detect ten different \emph{Arduino} programs which are labelled from $0$ to $9$.}
\label{fig:arduino-confusion-matrix}
\end{figure}

\textbf{Results:} Figure~\ref{fig:arduino-confusion-matrix} illustrates the confusion matrix of the classification results. The programs subject to the experiment are labelled from $0$ to $9$ in the figure. As can be seen, the majority of the Arduino programs were detected by the classifier accurately. Under a $10$-fold cross-validation, the classifier achieved a mean classification accuracy of $90$\% for an error margin of $11$\% within a $95$\% confidence interval. Considering the fact that currently it is nearly impossible to identify the software activities of an IoT device without a significant support from the manufacturer, the achieved accuracy through EM-SCA can potentially be a significant benefit to an investigator to gain insight on the device.

%Table~\ref{tab:arduino-classification-results} shows the classification results of the four software activities with slight changes from one another.

%\begin{table}[!htbp]
%\centering
%\begin{tabular}{|c|c|c|c|}
%\hline
%\rowcolor[HTML]{DAE8FC} 
%\textbf{Activity} & \textbf{Precision} & \textbf{Recall} & \textbf{F1-Score} \\ \hline
%\textbf{loop 1}    & 1.00               & 0.99            & 0.99              \\ \hline
%\textbf{loop 2}  & 1.00               & 0.99            & 1.00              \\ \hline
%\textbf{loop 3}  & 1.00               & 1.00            & 1.00              \\ \hline
%\textbf{loop 4}     & 0.98               & 1.00            & 0.99              \\ \hline
%\end{tabular}
%\caption{Arduino activity detection results. \textbf{Instead of a table, plot a colour coded confusion matrix for the 10 class classification.}}
%\label{tab:arduino-classification-results}
%\end{table}

\subsection{Detecting Modifications to Firmware}

A very simple IoT device with a $8$-bit processor and few kilobytes of memory is only capable of running a simple firmware that can perform a simple and repetitive task. The firmware running on such IoT devices are easier to be replaced by attackers in order to make them run malicious code. A device with a modified firmware can cause malfunctions not intended by the manufacturer. For example, \emph{Mirai} is a malware that infected certain types of IoT devices through exploiting their unchanged factory default passwords~\cite{antonakakis2017understanding}. It enabled the infected IoT devices to take part in \emph{distributed denial of service} (DDoS) attacks without the knowledge of device owner. Therefore, detecting such modifications to the stock firmware of an IoT device is highly necessary.
When the EM emission signature of a target device is already known, any change to the default firmware of the device should cause a detectable change to the EM emission pattern. Therefore, it is possible to train a machine learning model to recognize anomalous EM emission patterns due to firmware changes.

When detecting anomalies using machine learning models, there are two potential directions; namely outlier detection and novelty detection. In outlier detection, an unsupervised approach is taken where both legitimate data and anomalous data are provided to the machine learning model. The model fits into the legitimate data with the assumption that this data are densely packed in the space while anomalies stay comparatively away. In contrast, novelty detection is a semi-supervised approach where only the legitimate data samples are provided to the model to train. Whenever new data samples are provided, the model assesses the likeness of the new data to the data it was trained on in order to determine whether the new data belongs to the same distribution or not.

Since there are infinite possibilities for modification to the default firmware of an IoT device, it is difficult to provide sufficiently representative set of samples of anomalies for a machine learning model to learn. Therefore, in this case, semi-supervised novelty detection by training a model with only the legitimate samples is decided the best technique. In this experiment, a \emph{one-class SVM} with a non-linear kernel (RBF) provided by the \emph{scikit-learn} library was used for this purpose~\cite{scikit-learn}. When training the model, one of the Arduino programs used in the aforementioned software behaviour detection experiment was used as the legitimate firmware of the device, while a mixture of other programs were used as the modified programs. The model was trained by providing $500$ training samples of the legitimate program produced during the previous experiment. For testing, $100$ further samples of the legitimate program was provided where the testing error rate was $18$\%. Finally, when $20$ different modified Arduino programs were provided for validation, each of them were detected by the model recording a $100$\% accuracy on anomalous program detection.

%Training error rate: 36/500
%Testing error rate: 18/100
%Outlier error rate: 0/20

%\subsection{Effects of the Signal Sampling Rate}
\subsection{Challenge of Evidence Data Storage}

When listening to radio frequency data with a SDR, extremely large sampling rates are used to increase the amount of information captured. When this data are saved into files, the EM trace file sizes are considerably large even for small time windows. For example, consider a scenario where a HackRF SDR is capturing EM data on $20$~MHz sampling rate for a period of $1$~minute. Each sample generated by the HackRF device through \emph{GNURadio Companion} software consists of two $32$~bit float values representing \emph{Quadrature} and \emph{In-phase} components of the sample in I-Q interleaved stream format. This means, each I-Q sample is $8$~bytes long. Therefore, the size of the $1$~minute signal capture is approximately $9$~GB ($8$~bytes $\times$ $20$~MHz $\times$ $60$~seconds $\approx$ $8.94$~GB). In order to apply EM-SCA techniques and machine learning algorithms, thousands of such EM traces are required - making the management of data extremely challenging.

When capturing unintentional EM emissions from target IoT devices, it is necessary to capture a large enough bandwidth surrounding the target EM frequency. This is due to the fact that multiple side-band peaks that occur around the center frequency can contribute to the distinguishing between two different signals in the frequency domain. However, due to the inherent nature of software defined radio devices, the \emph{bandwidth} and the \emph{sampling rate} of signal capture are interdependent and it is impossible to reduce the sampling rate of the device without reducing the bandwidth. The contradicting requirements of reducing the sampling rate in order to reduce the file sizes and increasing sampling bandwidth in order to increase the amount of frequency domain information captured, necessitates the addition of an extra layer of processing of the captured data before saving. This can be done by collecting data with the highest bandwidth possible with the SDR device and then \emph{down-sampling} the data before saving into EM trace files. However, the question arises whether such down-sampling affects the signal classification accuracy when such low sample rate data are used with machine learning models.

In order to evaluate the correlation between sample rate of the signal acquisition device and the classification accuracy of machine learning models, the following experiment was performed. EM traces were captured for $4$~different Arduino programs with a sampling rate of $20$~MHz. Similar to the previous experiments, the information leaking center frequency of the target device was selected as $288$~MHz. After capturing $600$~EM traces per class, each of the trace files were \emph{down-sampled} in order to create new sets of EM trace files that has various sampling rates; namely $16$, $12$, $8$, $4$, $3$, $2$, $1$, and $0.5$~MHz. Using each data set (representing its unique sample rate) a Neural Network-based classifier was trained and tested. After performing a $10$-fold cross-validation for each classifier, the average \emph{F1-score} was taken along with the $95$\% confidence interval.

\begin{figure}[!t] %[!htbp]
\centering
\includegraphics[width=0.5\textwidth]{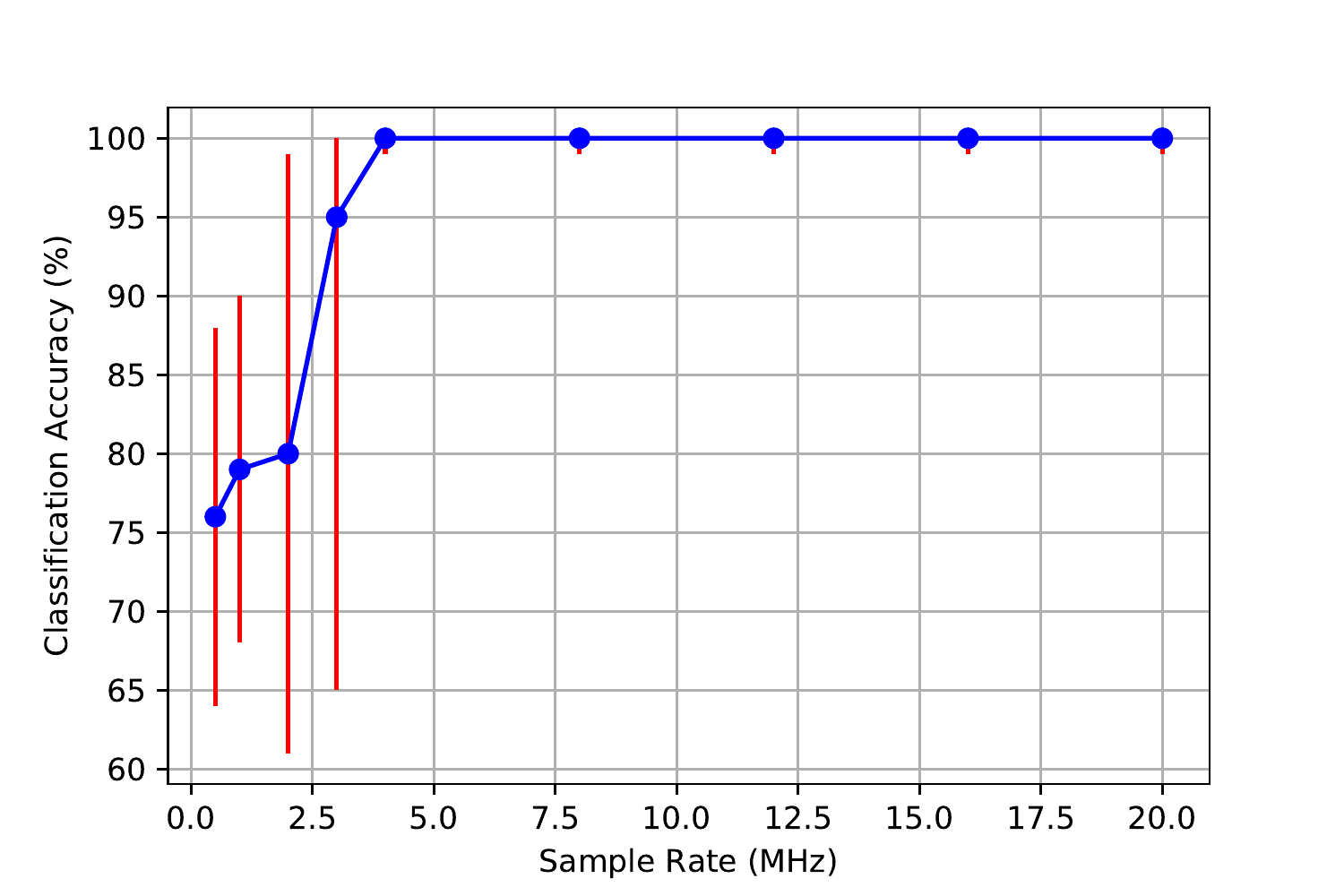}
\caption{The effect of EM trace sample rate to the signal classification accuracy when used with $4$~class classifier to identify four different \emph{Arduino} programs.}
\label{fig:sampling-rate-vs-classification-accuracy}
\end{figure}

Figure~\ref{fig:sampling-rate-vs-classification-accuracy} illustrates the variation of classification accuracy against the sampling rate where it is evident that the classification accuracy is not affected by sampling rates as low as $4$~MHz. However, when the sample rate goes below $4$~MHz, the classification accuracy plummets along with a significant increase in the classification error margin, depicted in red in Figure~\ref{fig:sampling-rate-vs-classification-accuracy}. This result indicates that it is possible to keep the bandwidth of the SDR at the maximum possible value, while down-sampling the data before saving into EM trace files without negatively affecting the performance of classification algorithms. Considering the maximum sample rate of the HackRF, i.e., $20$~MHz, and the lowest possible sample rate that did not adversely affect the classification accuracy in this experiment, i.e., $4$~MHz, it is possible to save $80$\% of the previously required space to store the EM trace data files. This could be a significant advantage when capturing EM data in on-site usage scenarios with portable equipment.

\subsection{Signal Analysis in Real-time}

\begin{figure}[!t] %[!htbp]
\centering
\includegraphics[width=0.5\textwidth]{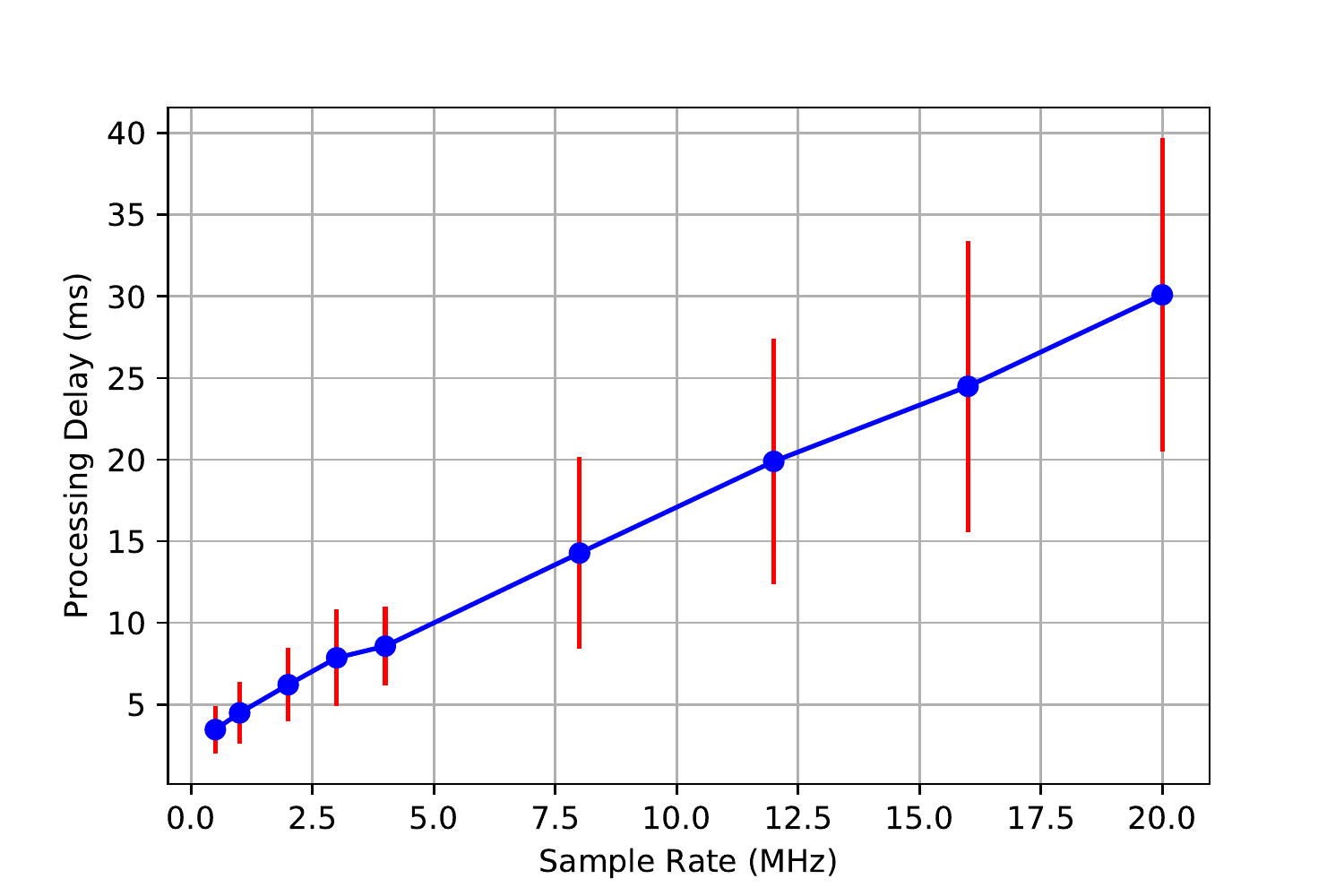}
\caption{The variation of EM data processing overhead against sampling rate when used with $4$~class classifier to identify four different \emph{Arduino} programs.}
\label{fig:sampling-rate-vs-processing-overhead}
\end{figure}

\begin{figure*}
    \centering
    \includegraphics[width=0.85\textwidth]{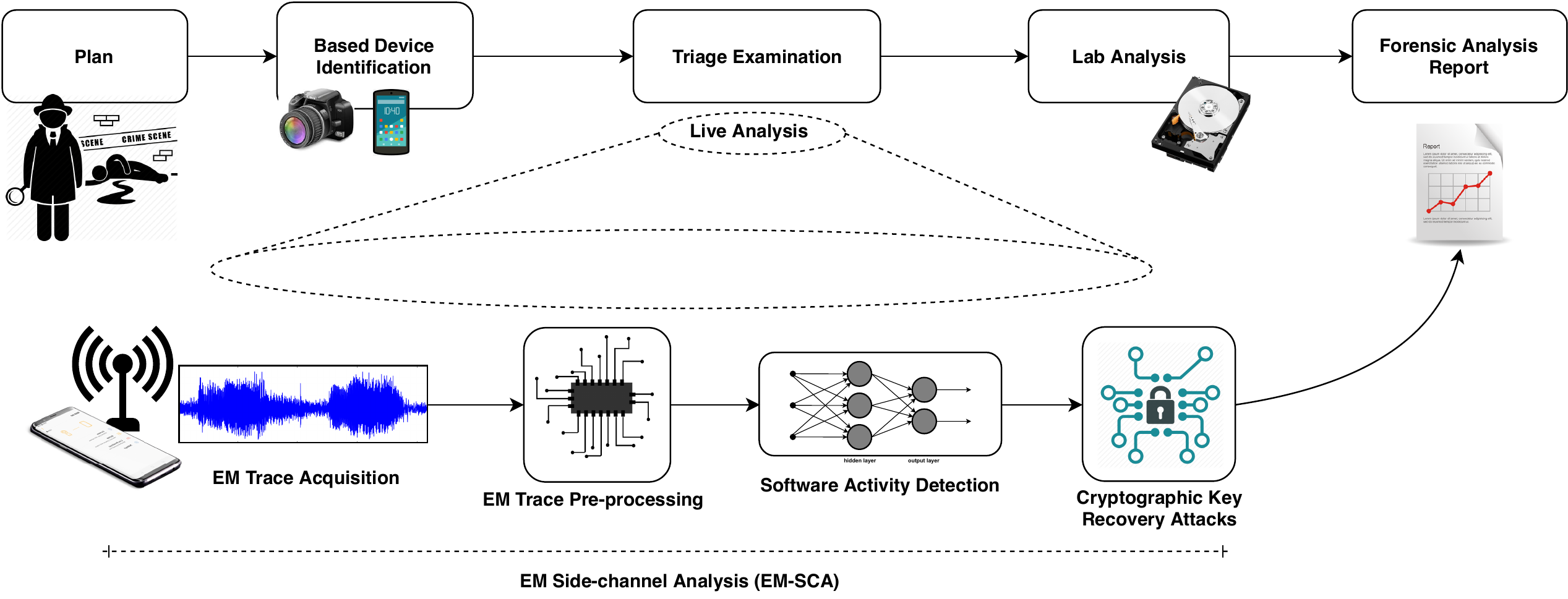}
    \caption{Integration of EM-SCA techniques into the standard digital evidence acquisition workflow (Adapted from Du et al.~\cite{du2017processmodelsdfaas}).}
    \label{fig:em-forensic-process}
\end{figure*}

Initial experimentation used EM trace data captured and saved into I-Q interleaved data files, which were later processed and used to train several machine learning classifiers. However, when using such machine learning-assisted EM-SCA for the live forensic analysis of IoT devices, real-time analysis is necessary. This is a challenging task since data preprocessing and classification tasks have to be performed within a tight time window in order to keep up with the real-time I-Q data stream.

When delivering EM samples in real-time from SDR devices to multiple software applications \emph{Transmission Control Protocol} (TCP) sockets are commonly used.
%For example, \emph{GNURadio} library which facilitates acquiring data from SDR devices makes use of the \emph{ZeroMQ} messaging protocol which is built on top of TCP.
Therefore, in order to maintain a stable real-time data processing system, the data preprocessing and machine learning stages must perform faster than the TCP retransmission timeout. On a Linux system, this timeout is typically set to $200$~ms and incremented at each timeout up to $15$ times. Figure~\ref{fig:sampling-rate-vs-processing-overhead} illustrates the variation of the processing delay of captured data windows against the sampling rate of the SDR device. Even at the highest sampling rate of the HackRF SDR, i.e., $20$~MHz, the processing delay does not exceed $40$~ms, which is well below the TCP retransmission timeout.

\section{Incorporation into the Digital Forensic Workflow}
%\section{Discussion}
\label{incorporating-EM-SCA-digital-forensics}
%%%%%%%%%%%%%%%%%%%%%%%%%%%%%%%%%%%%%%%%%%%%%%%%%%%%%%%%%

When a computing device is subject to a digital forensic investigation, the major focus is directed towards the non-volatile storage of the device, such as hard disks and removable media. Furthermore, in some cases, the analysis of the volatile memory of the device can be necessary as well. For example, if the device is running at the time it was seized, live data forensics can recover critical information such as temporary application data and cryptographic keys~\cite{Sayakkara:2018:ESA:3236454.3236512}. However, this triage examination of a device is still a highly unreliable task compared to analyzing disk images~\cite{du2017processmodelsdfaas}.

IoT devices, by definition, are always connected to a network enabling them to deliver their data to outside servers. Therefore, most IoT devices store a minimal amount of data on-board. When an IoT device data is vital to make progress in an investigation, the data has to be acquired from the cloud back-end where the IoT data is delivered to. The reliability of this IoT related evidence depends on the reliability of the IoT device's behaviour which is difficult to asses. This is where the EM-SCA techniques can help a digital forensic investigator to quickly assess an IoT device.

Figure~\ref{fig:em-forensic-process} illustrates where the EM-SCA has to fit into the current digital forensic analysis workflow. As IoT devices are usually designed to be always on, it is highly possible to have a seized IoT device still operational. Before attempting to acquire any non-volatile storage data physically from the device, the device must be switched off in order to prevent any physical damage to the data. The longer the device is running, the higher the risk of contaminating device data. If the device is connected to the network, it can potentially receive remote commands to wipe its internal storage. This situation highlights the necessity of performing EM-SCA within as short a time window as possible. The shorter the time window available to observe the EM emissions from a target IoT device, the smaller the amount of EM traces that can be used in the analysis algorithms. This poses a significant challenge for EM-SCA attacks on IoT devices under practical circumstances.

This work demonstrated the possibility of building machine learning models that are trained to detect a specific software related activity of a specific target IoT device. Detection of modifications to the stock firmware of a device and determining the cryptographic and non-cryptographic software activities in real-time can be useful for a forensic investigator to get an insight on a device. As Arduino and Raspberry Pi are representative of the two ends of the IoT device ecosystem in terms of computational resources, the results indicates that it should be possible to apply such machine learning based EM-SCA approaches to any IoT device.
Further research is necessary to explore the methods of building generalized machine learning models to cover commonly encountered IoT devices in digital investigations. Similarly, it is important to perform the evaluations with the stock firmware of such devices, especially for detecting malicious software modifications. It is necessary to make such machine learning-based EM-SCA to generate reliable information with a minimal amount of EM emission observations.

Legal investigations require a significant amount of reliability for digital evidence to be admissible in a court of law without reasonable doubt. The capabilities demonstrated with EM-SCA combined with the machine learning approach described in this work need to be time-tested before being used as a reliable evidence source. However, EM-SCA techniques can provide helpful directions for an investigator in order to uncover admissible evidence. Furthermore, development of side-channel mitigation techniques is an on-going interest for computer security researchers. Of course, this can pose a threat to digital forensic investigation with EM-SCA. However, the firmware of many IoT devices are not updated after they are shipped. Therefore, a known EM side-channel of an IoT device can remain exploitable for the rest of its service life~\cite{SAYAKKARA2019}. 

\section{Conclusion}
\label{conclusion}
%%%%%%%%%%%%%%%%%%%%%%%%%%%%%%%%%%%%%%%%%%%%%%%%%%%%%%%%%

As modern digital forensic investigations are increasingly encountering IoT data sources that provide vital information to solve cases, the need for non-intrusive and reliable ways of inspecting IoT devices arises strongly. This research work highlighted the potential of EM-SCA techniques combined with machine learning algorithms to tackle this problem. Using two representative IoT devices, a series of experiments were performed to demonstrate that the internal activities of IoT devices can be identified with a significant reliability. An attempt to classify cryptographic algorithms running on a high-end IoT device indicated that over $82$\% classification accuracy can be achieved with a very simple neural network-based classifier in a $4$~class classification problem. Similarly, a variety of similar programs, running on a low-end IoT device were detected over $90$\% accuracy indicating that, EM-SCA can be employed to distinguish between different software activities of IoT devices with a very detailed granularity. The same approach was demonstrated to be highly successful in a binary classification problem in order to detect whether the stock firmware running on a device has been tampered or not with an impressive accuracy of $100$\%.

While EM-SCA based software activity detection is usable to gather forensically useful insights about an IoT device, it is highly necessary to perform such EM-SCA procedures in real-time at the point of seizure of an IoT device while the device is still in operation. The experimentation presented as part of this paper demonstrated that it is possible to reduce the rate of EM signal samples through \emph{down-sampling} without any significant effect on the classification accuracy of the machine learning algorithms. This enables digital forensic investigators to process EM data on portable computing devices with less computing resources to get forensically useful insights on-the-spot without waiting for the reports of offline forensic analysis. Furthermore, it was shown that EM data acquisition from a SDR device and real-time processing of them involving machine learning algorithms can be performed successfully indicating that practically usable digital forensic software tools can be implemented on top of such an infrastructure.

%%%%%%%%%%%%%%%%%%%%%%%%%%%%%%%%%%%%%%%%%%%%%%%%%%%%%%%%%
\subsection{Future Work}
\label{future}
%%%%%%%%%%%%%%%%%%%%%%%%%%%%%%%%%%%%%%%%%%%%%%%%%%%%%%%%%

The outcomes of this research lays the foundation for further exploration on the applicability of EM-SCA on IoT devices for digital forensic purposes. To the end, the following avenues have been identified for future work.

\begin{itemize}[noitemsep,topsep=0pt]
    \item Minimization of the need for bespoke machine learning models for each individual type of IoT device through the development of models targeting commonly used processors and components across many devices. 
    \item Precise detection of the time periods that enclose individual cryptographic operations. This can aid in efficient reduction of the keyspace of the cryptographic keys used to protect IoT on-board data storage.
    \item Development of a metadata storage format in order to enable the management of EM traces acquired under digital forensic investigation scenarios by augmenting existing standards, such as \emph{HDF5}~\cite{folk2011overview}, \emph{VITA-49}~\cite{cooklev2012vita}, and \emph{SigMF}~\cite{hilburn2018sigmf}.
    \item Implementation of a ready-to-use, extensible EM-SCA analysis software framework for digital forensic investigators for IoT device inspection.
\end{itemize}

\bibliographystyle{model6-num-names}
\bibliography{bibfile}

%% not want to use model6-num-names.bst.



%% \bibitem{key}...
%%


% \end{thebibliography}

\end{document}